\documentclass[useAMS,usenatbib]{mn2e}
\usepackage{graphicx}                    
\usepackage{amssymb}                    
\usepackage{color}                       
\usepackage{url}                         

\title[Anisotropy of the entire inertial range of solar wind turbulence]{Power and spectral index anisotropy of the entire inertial range of turbulence in the fast solar wind.}

\author[R. T. Wicks, T. S. Horbury, C. H. K. Chen, and A. A. Schekochihin]{R. T. Wicks\thanks{E-mail: r.wicks@imperial.ac.uk}$^{1}$, T. S. Horbury$^{1}$, C. H. K. Chen$^{1}$, and A. A. Schekochihin$^{2}$\\$^{1}$ Space and Atmospheric Physics Group, Imperial College London, London, SW7 2AZ, UK.\\$^{2}$ Rudolf Peierls Centre for Theoretical Physics, University of Oxford, Oxford, OX1 3NP, UK.}

\begin{document}
\date{}

\pagerange{\pageref{firstpage}--\pageref{lastpage}} \pubyear{2009}

\maketitle

\label{firstpage}

\begin{abstract}

We measure the power and spectral index anisotropy of high speed solar wind turbulence from scales larger than the outer scale down to the ion gyroscale, thus covering the entire inertial range. We show that the power and spectral indices at the outer scale of turbulence are approximately isotropic. The turbulent cascade causes the power anisotropy at smaller scales manifested by anisotropic scalings of the spectrum: close to $k^{-5/3}$ across and $k^{-2}$ along the local magnetic field, consistent with a critically balanced Alfv\'{e}nic turbulence. By using data at different radial distances from the Sun, we show that the width of the inertial range does not change with heliocentric distance and explain this by calculating the radial dependence of the ratio of the outer scale to the ion gyroscale. At the smallest scales of the inertial range, close to the ion gyroscale, we find an enhancement of power parallel to the magnetic field direction coincident with a decrease in the perpendicular power. This is most likely related to energy injection by ion kinetic modes such as the firehose instability and also marks the beginning of the dissipation range of solar wind turbulence.

\end{abstract}

%

\begin{keywords}
turbulence -- plasmas -- solar wind.
\end{keywords}

\section{Introduction}

The solar magnetic field embedded in the turbulent solar wind \citep{Goldstein95} causes several kinds of anisotropy. Different physical processes operate parallel and perpendicular to this mean field. The fluctuations (with respect to the large-scale mean field) of the parallel and perpendicular components of the magnetic field scale differently from large \citep{Nicol09} to small scales \citep{Alexandrova, Sahraoui}. Many different solar wind quantities, for example; cross helicity \citep{Milano04}, the magnetic field correlation function \citep{Dasso05, Osman07, Weygand09}, and velocity components \citep{ChapmanHnat07}, have been shown to vary anisotropically with respect to the mean magnetic field direction.
\par
Recent advances in the study of the solar wind have shown that the local magnetic field direction causes spatial anisotropy in the statistical properties of the fluctuating fields along and across the local mean magnetic field. This anisotropy can be measured in correlation functions \citep{Matthaeus94}, power at a fixed scale \citep{Bieber96, Chen10}, and variation of the spectral index depending on the angle to the local mean field \citep{Horbury08, Podesta09, Chen10}; see \citet{Chen09} for a discussion of the relationship between these manifestations of the anisotropy. These results have shown that the measured spacecraft power spectra of the turbulent magnetic field, $\textbf{B}$, at different angles, $\theta_B$, to the local mean magnetic field are consistent with those predicted by the phenomenological theory of \citet{GS95} based on the conjecture of critical balance between linear (Alfv\'{e}n) and nonlinear timescales in the turbulent cascade. Thus far these studies have focused on the inertial range, the small-scale boundary of which is defined by the ion gyroradius $\rho_i$, and have not approached the large-scale limit, the outer scale, the point at which the power spectrum $k^{\alpha}$ rolls over to a flatter index of $\alpha \approx -1$ \citep[e.g.][]{Horbury96}.
\par
Here we extend the wavelet-based local magnetic field analysis of \citet{Horbury08} and \citet{Podesta09} to larger scales, including the outer scale. We investigate the properties of the turbulence, in particular the anisotropy, across the entire inertial range and demonstrate for the first time the development of the turbulent cascade from its original, nearly isotropic, state at low frequencies all the way down to to its final, anisotropic, state near the ion gyroscale.

\section{Wavelets and the Local Mean Magnetic Field}

We use the method devised by \citet{Horbury08} and explained in more detail by \citet{Podesta09}. The Morlet wavelet is used to decompose the time series of magnetic field observations into powers at given frequencies and scales. Neighbouring wavelet scales differ by a factor of $1.6$, making the Morlet wavelet close to orthogonal and ensuring a good coverage of frequency without oversampling. The wavelet scale is used as the length over which to calculate the mean magnetic field direction (the local mean field method). The power $P(f, \theta_B)$ calculated using the wavelet is then assigned to a bin corresponding to an angle $\theta_B$, the angle between the mean field and the sampling direction (the direction of the spacecraft flight in the solar wind) and a Fourier frequency $f$ associated with the wavelet scale \citep{Horbury08, Podesta09}. We adopt the Taylor hypothesis \citep{Taylor} that the flow speed is much greater than the sound and Alfv\'{e}n speeds and so a time series can be considered to be a one-dimensional cut through a time stationary plasma. Therefore, our observations are necessarily of the reduced spectrum \citep{FredricksCoroniti76}.
\par
We take periods of continuous Ulysses data from 1995 while the spacecraft is in fast polar solar wind. We use one second resolution magnetic field data \citep{Balogh}. Analysing each period produces a power spectrum ranging in spacecraft frequency $f$ between $3.3 \times 10^{-5}$ and $2.5 \times 10^{-1}$ Hz. We resolve angles $\theta_B$ between the observed wavevector and the local mean magnetic field in $10^{\circ}$ bins between $0^{\circ}$ and $180^{\circ}$. In what follows we only plot the bins $0^{\circ} \leq \theta_B < 90^{\circ}$: angles greater than $90^{\circ}$ do not occur frequently and so we cannot reliably measure the anti-parallel power spectrum.
\par
Ulysses observations are made at $1$ s cadence, however, the important physical scale for kinetic plasma physics in the solar wind is the proton gyroscale $\rho_i$. In order to compare different periods directly and to cast our results in physically relevant units we convert the spacecraft observation frequency into a flow-parallel wavenumber $k$ by dividing by the average solar wind speed $|V|$ and normalise this by $\rho_i$:
\begin{equation}
k\rho_i = \frac{2 \pi f \rho_i}{|V|} = \frac{2 \pi f \sqrt{2 k_B T_i m_i}}{e |V| |B|} ,
\label{eq:1}
\end{equation}
where $k_B$ is Boltzmann's constant, $T_i$ is the proton temperature, $m_i$ is the mass of a proton, $e$ is the charge on an electron and $|B|$ is the magnetic field strength. 
\par
\begin{figure}%
\includegraphics[width=\columnwidth]{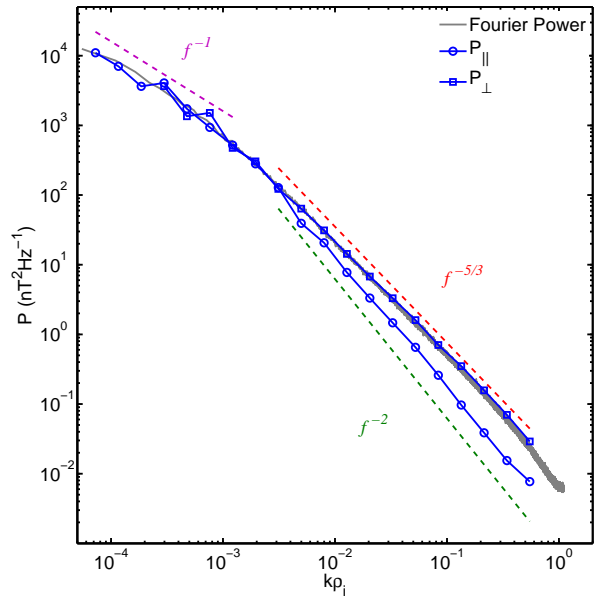}%
\caption{Trace of the wavelet and Fourier power spectra of magnetic field observations from Ulysses for the period between days 100 and 200 of 1995. Frequencies are converted to wavenumbers using the solar wind velocity and normalised to the ion gyroradius $\rho_i$ (Equation \ref{eq:1}). See Figure \ref{fig:4} for compensated spectra.}
\label{fig:1}%
\end{figure}

\section{Anisotropy of the Entire Inertial Range}

First, we analyse a period of fast polar wind from Ulysses data between days 100 and 200 of 1995; during this time Ulysses moved from a solar latitude of $28^{\circ}$ to $79^{\circ}$ and from a heliocentric distance of $1.38$ AU to $1.93$ AU. This long interval provides an accurate anisotropic power spectrum at the lowest frequencies used here; shorter periods can be used if angular resolution is not required at such low frequencies. Figure \ref{fig:1} shows the trace of the magnetic field power tensor, averaged over periods when the solar wind flow is parallel, $P_{||}$ ($0^{\circ} \leq \theta_B < 10^{\circ}$) and perpendicular, $P_{\perp}$ ($80^{\circ} \leq \theta_B < 90^{\circ}$) to the local magnetic field calculated using wavelets and the average windowed Fourier power calculated between the same frequencies for the same period. At the smallest values of $k\rho_i$ the power is isotropic and all three lines lie close together with a spectral index of approximately $-1$. At $k\rho_i \approx 3\times10^{-3}$, $P_{||}$ begins to diverge from the Fourier power and $P_{\perp}$. The power anisotropy increases as $k\rho_i$ increases from this point; $P_{\perp}$ and the Fourier power follow each other very closely and are a factor of five larger than $P_{||}$ at the largest $k\rho_i$ measured. We stress that the use of wavelets to analyse the anisotropy of the magnetic field means that the magnetic field is not broken into components parallel and perpendicular to the field. Thus the terms $P_{||}$ and $P_{\perp}$ do not refer to components of the field but to the mean trace power in the field when the flow past the spacecraft is parallel or perpendicular to the mean field.
\par
\begin{figure}%
\includegraphics[width=\columnwidth]{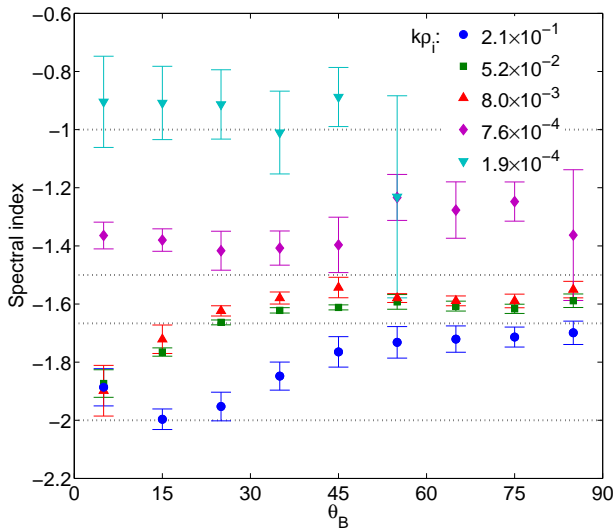}
\caption{Dependence of the spectral index in different scale ranges on angle to the local mean magnetic field direction.}
\label{fig:2}%
\end{figure}
In Figure \ref{fig:2} we show the dependence of spectral index $\alpha$ on scale by plotting $\alpha$ determined by a least-squares fitting in logarithmic space of a straight line to power measurements. Each fit uses five wavelet power measurements and so covers scales separated by a factor of $10.5$. The $k\rho_i$ shown are the central values for each five point window, the uncertainties are calculated as the error from the fitting of the line. At large scales ($k\rho_i = 1.9\times10^{-4}$), the power spectra have an isotropic spectral index $\alpha \approx -1$. Above $k\rho_i \approx 5\times10^{-4}$, the spectrum steepens (diamonds in Figure \ref{fig:2}). As $k\rho_i$ increases, the spectral indices decrease further and their values vary with $\theta_B$ but only slowly with scale (squares and triangles in Figure \ref{fig:2}). These scales show a similar result to those of \citet{Horbury08} and \citet{Podesta09} and confirm that in the inertial range fluctuations have a parallel spectral index of $-2$, while the perpendicular spectral index lies between the $-5/3$ of a critically balanced Alfv\'{e}nic cascade \citep{GS95} and the $-3/2$ predicted by the theory of dynamically aligned Alfv\'{e}nic turbulence \citep{Boldyrev06}. We will show in the next section that the data, in general, support a value closer to $-5/3$.
\par
The two decade wide range of scales between $k\rho_i = 3\times10^{-3}$ and $k\rho_i = 1\times10^{-1}$ over which the spectral indices vary only slightly can be considered by definition to be the inertial range. At the largest values of $k\rho_i$ (blue circles in Figure \ref{fig:2}), the spectral index decreases compared to the inertial range for all $\theta_B > 10^{\circ}$ and a value close to $-2$ is found for a wider range of $\theta_B$. This is the beginning of the effect of kinetic plasma physics close to the ion gyroscale \citep[e.g.][]{Schekochihin09}, as we show in the next section. Thus, we have covered the entire inertial range of MHD turbulence, from the effective outer scale defined by the break point from $1/f$ scaling at the lowest frequencies, to the beginning of kinetic effects at the ion gyroscale. 

\section{Dependence on Heliocentric Distance}
\begin{table*}
\begin{tabular}{ccccccccccc}
\hline
Start&R&$|B|$&$V_{SW}$&$V_A$&$n_i$&$\sigma_c$&$T_i$&$\rho_i$&\multicolumn{2}{c}{Inertial Range $\alpha$} \\
Day&(AU)&(nT)&(km s$^{-1}$)&(km s$^{-1}$)&(cm$^{-3})$& &($\times10^5$K)&($\times10^3$km)&$\alpha(P_{\perp})$&$\alpha(P_{||})$\\
\hline
$100$&$1.48$&$2.82$&$758$&$56$&$1.21$&$0.65$&$2.49$&$1.52$&$-1.62\pm0.03$&$-1.92\pm0.02$\\
$150$&$1.76$&$1.98$&$780$&$49$&$0.77$&$0.66$&$2.09$&$1.98$&$-1.58\pm0.02$&$-2.00\pm0.01$\\
$200$&$2.11$&$1.46$&$785$&$45$&$0.50$&$0.57$&$1.91$&$2.57$&$-1.64\pm0.01$&$-1.9\pm0.1$\\
$250$&$2.46$&$1.17$&$779$&$41$&$0.38$&$0.52$&$1.75$&$3.06$&$-1.69\pm0.01$&$-1.9\pm0.1$\\
$300$&$2.77$&$0.97$&$766$&$38$&$0.31$&$0.46$&$1.61$&$3.56$&$-1.64\pm0.02$&$-1.94\pm0.04$\\
\hline
\end{tabular}
\caption{Results from five different periods of Ulysses data from 1995, showing start day and mean physical parameters for each period. The spectral indices of the parallel and perpendicular inertial range are calculated fitting straight lines to the logarithm of the spectra in the range $2\times10^{-2} < k\rho_i < 2\times10^{-1}$.}
\label{table:1}
\end{table*}
\par
As the solar wind travels away from the Sun, the ion gyroscale increases in size, the mean magnetic field strength decreases, the collisional and turbulent age of the plasma increases, and the outer scale moves to lower frequencies. Here we perform the same analysis as in the previous section for five 50 day periods of Ulysses data from 1995, taken at different distances from the Sun and so investigate the dependence on heliocentric distance of the anisotropy and spectral break points.
\par
A summary of mean plasma parameters for each period is given in Table \ref{table:1} as well as the anisotropic spectral indices and the mean of the normalised cross helicity, $\sigma_c = (2\textbf{v} \cdot \textbf{b})/(\textbf{v}^2 + \textbf{b}^2)$ where \textbf{v} and \textbf{b} are fluctuations of the velocity and magnetic field in Alfv\'{e}n units; $\sigma_c$ is calculated spectrally for the whole of each period and the values quoted here are the mean values between $6\times10^{-5}$ and $8\times10^{-5}$ Hz, which corresponds to the lowest frequency wavelet used. The values of $\sigma_c$ are consistently above zero, showing the imbalanced nature of the turbulence. The values of the spectral indices in Table \ref{table:1}, calculated in a fixed range of scales $2\times10^{-2} \leq k\rho_i \leq 2\times10^{-1}$ for all five periods, are similar and broadly consistent with $-5/3$ and $-2$ found in the previous section. Due to the changing gyroradius and magnitude of \textbf{B}, this would not be true if we used a fixed range of frequencies, we would instead see increasingly negative values of the spectral index as the solar wind moves away from the Sun. This is because as the ion gyroradius increases with distance from the Sun a set frequency can become comparable to, or smaller than the ion gyroscale, and thus slip into the dissipation range, where the spectra are steeper \citep{Chen10}.
\par
\begin{figure}%
\includegraphics[width=\columnwidth]{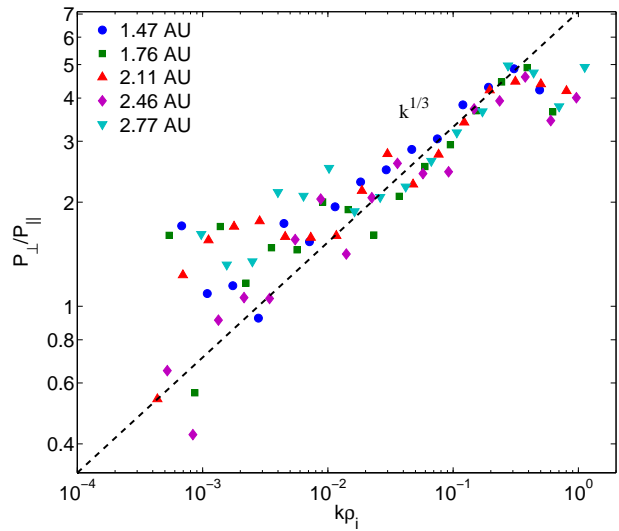}
\caption{Dependence of power anisotropy for each of the five periods in Table \ref{table:1} on wavenumber scaled to the ion gyroscale, $k\rho_i$.}%
\label{fig:3}%
\end{figure}

Figure \ref{fig:3} shows the power anisotropy, $P_{\perp}/P_{||}$, a quantity first introduced by \citet{Podesta09}, as a function of $k\rho_i$ for the five periods. The data from different periods collapse onto a single line, particularly at larger wavenumbers. Its slope at scales within the inertial range is approximately $1/3$, which is the value expected for a critically balanced turbulence of \citet{GS95}, because $k^{-5/3}/k^{-2} = k^{1/3}$. There is a peak at $k\rho_i = 0.4$ and a trough at $k\rho_i = 0.7$, consistent with Figure 6 of \citet{Podesta09}. For $k\rho_i < 10^{-2}$ the values of $P_{\perp}/P_{||}$ mostly rise above the $k^{1/3}$ line and fill the region $1 < P_{\perp}/P_{||} < 2$ which is the region of near-isotropy corresponding to the roll over to a spectral index of $-1$ shown in Figure \ref{fig:1} for the closest period to the Sun.
\par
All five periods have a similar peak power anisotropy of $P_{\perp}/P_{||} \approx 5$. This is close to the value of $7$ reported by \citet{Podesta09} for power anisotropy in the fast wind in the ecliptic at $1$ AU. We note that \citet{Podesta09} used smaller angular bins than we have here, which could decrease the measured power in the $P_{||}$ angle bin, thus increasing the anisotropy (there is likely to be little difference in $P_{\perp}$ due to the approximate invariance of power at angles close to $90^{\circ}$). 
\par
We see that the peak power anisotropies of all five periods are very similar. If we assume a universal anisotropic scaling in the inertial range and take the result shown above that the outer scale is nearly isotropic, these results imply, perhaps surprisingly, that the total width of the inertial range does not change significantly with distance from the Sun. Indeed Figure \ref{fig:3} shows that the range of scales from where $P_{\perp}/P_{||} = 1$ to $k\rho_i = 1$ is always approximately two decades ($5\times10^{-3} \lesssim k\rho_i \lesssim 5\times10^{-1}$). While (as previous studies have shown \citep{Bavassano82, Horbury96} both the outer scale and the ion gyroscale increase with distance from the Sun (Table \ref{table:1}), they increase in such a way as to keep the total width of the inertial range approximately constant. 
\par
Let us make a simple estimate of how the ratio of $\rho_i$ and the outer scale ($L$) varies with distance from the Sun $R$. We assume $|B| \propto R^{-1.48}$, $|V| \approx$ constant, $T \propto R^{-1.02}$ and $L \propto R^{1.1}$, the scalings that have been obtained from Ulysses observations \citep{Ebert09, Goldstein96, Horbury96}. Then:
\begin{equation}
\frac{L}{\rho_i} \propto \frac{L|B||V|}{\sqrt{T}} \propto R^{0.13} .
\label{eq:3}
\end{equation}
This very weak dependence on $R$ is essentially unmeasurable due to the scatter in the power anisotropy measurements at large scales and the small range of heliocentric distances covered. This explains why we do not see a significant increase in the width of the inertial range. 
\par
This simple analysis taken in conjunction with the scaling of the magnetic field strength from the Parker spiral equation also leads to the conjecture that the inertial range might be expected to be wider closer to the Sun. Close to the Sun the magnetic field decreases like $|B| \propto R^{-2}$, whereas further out in the heliosphere it decreases with $|B| \propto R^{-1}$ \citep{Burlaga84, Burlaga02}. This would imply $L/\rho_i \propto R^{-1/2}$ close to the Sun and $L/\rho_i \propto R^{1/2}$ further out in the heliosphere. The Ulysses results appear to be in the transition region between these two behaviours where the $L/\rho_i \sim const$. Thus, the inertial range in the corona could be wider than that observed in fast solar wind at 1 AU and wider again in the outer heliosphere. The winding of the Parker spiral controls the scaling of the magnetic field magnitude with radius and so this scaling is also dependent on heliospheric latitude.
\par
\begin{figure}%
\includegraphics[width=\columnwidth]{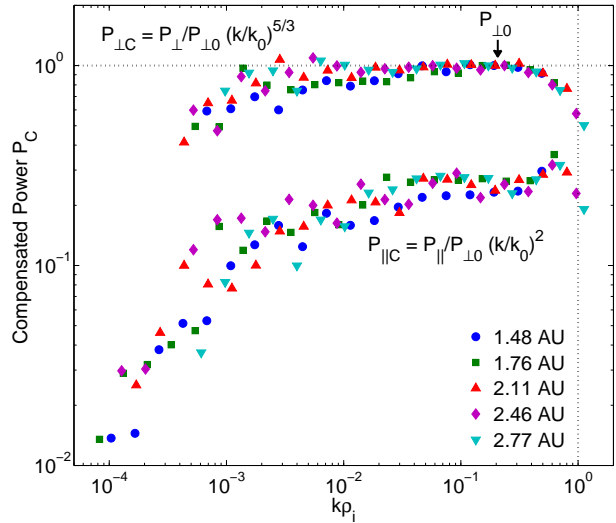}
\caption{Perpendicular and parallel power for each of the five periods in Table \ref{table:1}, compensated to remove a spectral gradient of $-5/3$ from the perpendicular power and $-2$ from the parallel.}%
\label{fig:4}%
\end{figure}

Finally, we show the similarity between the five periods we analyse and the effect of power enhancement parallel to the field close to the ion gyroradius by plotting compensated spectra for the parallel and perpendicular power. Compensated spectra are defined:
\begin{equation}
P_C(k) = \frac{P(k)}{P_{\perp 0}} \left( \frac{k}{k_0} \right)^{-\alpha} ,
\label{eq:2}
\end{equation}
where $k_0\rho_i = 2\times10^{-1}$ and $P_{\perp 0} = P_{\perp}(k_0\rho_i)$; indicated in Figure \ref{fig:4} by an arrow. The spectral indices used to compensate the spectra are $\alpha = -5/3$ for the perpendicular spectrum and $\alpha = -2$ for the parallel spectrum. 
\par
Both $P_{\perp}$ and $P_{||}$ show a horizontal region between $2\times10^{-2} \lesssim k\rho_i \lesssim 5\times10^{-1}$ implying that the spectral indices are close to $-5/3$ and $-2$ for the perpendicular and parallel power, respectively. At $k\rho_i < 10^{-2}$ the parallel spectrum turns downward, the steeper gradient indicating the roll over towards isotropy and a spectral index of $-1$ has begun. The outer scale break point is at $k\rho_i \approx 10^{-3}$ where another downward turn is seen in both the perpendicular and parallel spectra. 
\par
At $k\rho_i \approx 0.7$, a clear peak can be seen in the parallel power which coincides with the steepening of the perpendicular spectrum. This is remarkably consistent across all five periods used. The parallel power peak suggests a local enhancement (perhaps injection) of energy possibly due to ion kinetic instabilities, but does not of course prove it. Ulysses high speed wind particle distributions are often close to the firehose instability threshold (\citealt{Matteini07}, see also simulations by \citealt{Hellinger08}). The enhancement in $P_{||}$ at $k\rho_i \approx 0.7$ is consistent with fluctuations whose wavevectors are parallel to the field, although the larger power when $\theta_B \approx 90^{\circ}$ means that such an enhancement would not be detectable in the perpendicular power, so the total wavevectors may be oblique. What is clear is that the anisotropy is reduced at this scale, perhaps by an admixture of fluctuations with $k_{||} \geq k_{\perp}$.
\par
Note that the peak in the parallel power is clearly in the same place for all spectra plotted using the gyroscale normalisation, despite the fact that frequencies and power levels change significantly over the 5 periods used. This means that what we see is a physical effect associated with the gyroscale and not an instrumental or spacecraft effect.

\section{Conclusions}
We have used a wavelet analysis of Ulysses fast polar solar wind observations to study the anisotropy of the entire inertial range of turbulence with respect to the local mean magnetic field. At the outer scale turbulence is isotropic. The spectral index is approximately $-1$ for scales larger than the outer scale and changes to $-5/3$ at smaller scales.
\par
At scales smaller than the outer scale, the turbulence develops a spatial anisotropy with smaller power in fluctuations that vary along the local mean magnetic field and a corresponding spectral index of $-2$, while the larger power in cross-field fluctuations has a $-5/3$ spectrum. These spectral scalings are consistent with a critically balanced Alfv\'{e}nic cascade \citep{GS95}. The perpendicular scaling appears closer to $-5/3$ than to $-3/2$ advocated by more recent theory \citep{Boldyrev06} and numerical simulations \citep[e.g.][]{Maron01}.
\par
The width of the inertial range remains constant at all heliocentric distances investigated. This can be understood by considering the change in the ratio of outer scale to the ion gyroscale with heliocentric distance and leads to the interesting conjecture that the inertial range is wider closer to the Sun and in the outer heliosphere (see Section 4).
\par
The small scale end of the inertial range seems naturally to scale with ion gyroradius. There appears to be an injection of power into parallel wavenumbers at $k\rho_i \sim 1$, conceivably due to the firehose instability.
\par
Our results show yet again that the solar wind is a unique laboratory for studying plasma turbulence across a wide range of scales in extraordinary detail. The isotropy at the outer scale and the remarkable consistency of the behaviour at the ion gyroscale show that solar wind turbulence, perhaps surprisingly, is quite a `clean' case and so is not only interesting for its own sake, but also as a representative case study of a magnetised plasma of the kind that is ubiquitous in the Universe.
\par
We would like to thank Miriam Forman, Petr Hellinger and Lorenzo Matteini for useful discussions. This work was funded by the STFC and by the Leverhulme Trust Network for Magnetized Plasma Turbulence.

\label{lastpage}

%
%


\end{document}